\documentclass{aa}
\usepackage{graphics}
\begin{document}
\thesaurus{}
\title{The Circumstellar Envelope of $\rm \pi^1$ Gru}

\author{G.R.\, Knapp\inst{1},
\and K. Young\inst{2} and M. Crosas\inst{2}}
\institute{Department of Astrophysical Sciences, Princeton University,
Princeton, NJ 08544, USA; gk@astro.princeton.edu
\and 
Harvard-Smithsonian Center for Astrophysics, 60 
Garden St., Cambridge, MA 02138, USA; rtm@dolson.harvard.edu,
mcrosas@cfa.harvard.edu}

\date{}

\titlerunning{$\rm \pi^1$ Gru}
\authorrunning{G.R. Knapp et al.}
\maketitle
\begin{abstract}

CO(J = 2--1) and SiO(J = 5--4) emission has been observed from the
molecular envelope around the nearby S star $\rm \pi^1$ Gru.
The CO line profile differs from the usual parabolic shape seen in
uniformly-expanding envelopes; it has a Voigt-like profile and two horns.
A model for line formation in the envelope shows that a tilted, expanding disk
reproduces the observations well.  
The star also has a fast molecular wind, with a projected
outflow speed of at least 70
and perhaps as high as 90
$\rm km~s^{-1}$. 
The fast wind is presumably ejected from the
poles of the disk.
These observations show that the complex structure seen in
many planetary nebulae, including quadrupolar structure and fast winds, may
largely evolve from structure formed while the progenitor star is in the
last stages of evolution on the AGB.
\end{abstract}

\section{Introduction}

Mass loss from stars on the Asymptotic Giant Branch (AGB) takes place as
a cool, dusty, molecular wind  at rates which are so high that they dominate
the evolution of the star.  The energy in these huge stars is produced 
in their very small cores, leading to the expectation that the structure
of the star and of its circumstellar outflow will be spherically symmetric.
However, a plethora of recent observations (e.g. Lucas \cite{lucas};
Monnier \cite{monnier}; Tuthill et al. \cite{tuthill}; Monnier et al. 
\cite{monnieretal})
have shown that the outflows are far from smooth,  steady, 
uniform and spherically
symmetric, and that asymmetries can be seen on a wide range of size scales.
This paper discusses  molecular line observations of the envelope of $\rm 
\pi^1$ Gru, which is known to have complex structure and large-scale 
asymmetry from previous CO
observations (Sahai \cite {sahai}, hereafter S92). 

$\rm \pi^1$ Gru is a nearby mass-losing S star, at a distance of 153 pc
(Perryman et al. \cite{perryman}; van Eck et al. \cite{eck}).
S92 found that the CO line
profiles have a very unusual shape: they are double-horned, and do not
have the steep sides typical of molecular line emission from uniformly
expanding stellar winds.  Further, the envelope has north-south kinematic
structure, and at least two velocity components; a ``normal'' outflow with
an apparent speed of about 11 $\rm  km~s^{-1}$, and a faster outflow with
a speed of $\rm \geq ~ 38 ~ km~s^{-1}$.

The molecular line observations measure a mean mass loss 
rate of several $\rm \times 10^{-7} ~ M_{\odot}~yr^{-1}$, while the
60$\rm \mu m$ emission from circumstellar dust shows
that the envelope is extended (Young et al. \cite{young}), with a 
radius of about $\rm 6 \times 10^{17}$ cm and a kinematic 
age of $\rm \sim 10^4$ years.  $\rm \pi^1$ Gru is the primary
of a wide binary system, whose secondary, at
a projected separation of $2.7''$, appears to be a solar-mass main
sequence star (Proust et al. \cite{proust}; Ake \& Johnston 
\cite{ake}).  The stars are
thus at least $\rm 6 \times 10^{15}$ cm apart and, since the system mass is 
at least 2 $\rm M_{\odot}$, the orbital period is about 6000 years and the
orbital velocity about 2 $\rm km~s^{-1}$.  

Given the possible presence of a fast molecular wind
and bipolar structure in the envelope of $\rm \pi^1$ Gru, 
plus the presence of its binary companion, 
we reobserved the envelope
with improved sensitivity and modeled its line emission
in an attempt to better define its geometry
and kinematics.  These observations are described in the 
next section, and the results are given in Section 3.  Section 4 describes
a model of the
structure of the circumstellar envelope.  These results are discussed in
Section 5, and Section 6
contains the conclusions.

\section{Observations}

The properties of $\rm \pi^1$ Gru from the literature, plus those derived
in the present paper, are summarized in Table 1.  The distance is from
the Hipparcos parallax (Perryman et al. \cite{perryman}), 
the luminosity from van Eck et al. (\cite{eck}),
and the variability type and period from Proust et al. (\cite{proust}),
and Kerschbaum \& Hron (\cite{kerschbaum}).

\begin{table}
\caption[]{Observed Properties of $\rm \pi^1 Gru$}
\vspace{0.5cm}
\begin{tabular}{lllll}
$\alpha$(1950)& $\rm 22^h ~19^m 41.13^s$\cr
$\delta$(1950)& $\rm -46^o ~ 12' ~ 02.4''$\cr
Spectral Type& S5.7\cr
Variable Type& SRb\cr
Period& $\rm 150^d$\cr
Distance& 153 +28, -20 pc\cr
$\rm L_{bol}$& 7450 $\rm L_{\odot}$\cr
& \cr
SiO(6--5):& \cr
I&  0.75 $\pm$ 0.05 $\rm K \times km~s^{-1}$\cr
$\rm T_{MB}$& 0.042 $\pm$ 0.005 K \cr
$\rm V_o$& 13.1 $\pm$ 2.1 $\rm km~s^{-1}$\cr
$\rm V_c$& -12.6 $\pm$ 0.8 $\rm km~s^{-1}$ \cr
& \cr
CO(2--1) (high resolution):& \cr
Projected $\rm V_o$& 5.1 $\rm km~s^{-1}$\cr
$\rm V_c $& -11.8 $\rm km~s^{-1}$\cr
Mean $\rm \mathaccent 95 M$& $\rm 4.5 \times 10^{-7} ~ M_{\odot} ~ yr^{-1}$\cr
& \cr
CO(2--1) (map):& \cr
Projected $\rm V_o$& 6.4 $\rm km~s^{-1}$\cr
$\rm V_c$& -11.7 $\rm km~s^{-1}$\cr
$\rm \mathaccent 95 M$& $\rm 1.2 \times 10^{-6} ~ M_{\odot}~yr^{-1}$\cr
$\rm V_o$& 15 - 18 $\rm km~s^{-1}$\cr
& \cr
\end{tabular}
\end{table}

The observations were made with the 10.4m Robert B. Leighton telescope of the 
Caltech Submillimeter Observatory on Mauna Kea, Hawai`i, on September
4-6 1997.  The
southerly declination of the star means that it is always at a low zenith
angle as observed from the CSO, and the consequent high air mass makes
submillimeter observations, which have higher angular resolution,
impractical.  The star was observed in 
the relatively transparent 1.2 mm window, in the CO(2--1) line at 
230.538 GHz and the SiO(6--5) line at 260.518 GHz.  The telescope 
half-power beamwidth was measured to be $\rm 30''$, and the main
beam efficiency to be 69\%, using observations of Jupiter. The 
zenith atmospheric opacity $\rm \tau_o$ was $\sim$0.06.

The observations used a liquid helium cooled SIS junction
receiver with a double sideband system temperature of $\sim$100 K.
The temperature scale and the 
atmospheric opacity were measured by comparison with a room
temperature load.  The temperature scale was corrected for the atmospheric
opacity and for the main beam efficiency.

The spectral lines were observed using two acousto-optic spectrographs
with bandwidths of 500 MHz and 50 MHz respectively over 1024 
channels.  The spectrometer frequency, frequency scale and spectral
resolution were calibrated using an internally generated frequency comb,
and the velocity scale was corrected to the Local Standard of Rest (LSR).
The velocity resolutions of the spectrometers were $\rm \sim
2.1 ~ km~s^{-1}$  and $\rm \sim 0.21 ~
km~s^{-1}$.

The observations were made by chopping between the star position and an 
adjacent sky position with the secondary mirror, using a chop 
throw of $\rm 120''$ in azimuth at a rate of 1 Hz.  Pairs of chopped 
observations were made with the star placed alternately in each beam.
The spectral baselines resulting from this procedure are linear to
within the r.m.s. noise for both spectrometers.

We performed three observations of the $\pi^1$ Gru envelope.  First, we
obtained a sensitive observation of the CO(2--1) line with both the
50 MHz (high resolution) and 500 MHz (large bandwidth)
spectrometers.  The high resolution 
observation was made to measure the detailed shape of the strong CO
line emission close to the systemic velocity of the star, while the large
bandwidth observation was made to measure the weak emission from
the star's high velocity molecular outflows.  During these observations,
the telescope pointing was checked roughly once per hour using the strong
line emission.  Second, we observed the SiO(6--5) line;
because this line requires high density for excitation, and because
much of the Si condenses onto dust close to the star, 
this emission arises from the 
inner regions of the envelope. 

The third observation was a map of the CO(2--1) emission from the
envelope made on a $\rm 100'' \times 100''$ square grid centered on the
star's position and sampled every $\rm 10''$.
The observation was made using the `on-the-fly' (OTF)
technique.  This method consists of
scanning across the object's position and accumulating spectra as the telescope
moves.  The grid is built up by a set of scans stepped in the orthogonal
direction. The map of $\rm \pi^1$ Gru was made by executing 
several rapid complete OTF maps and summing them; this procedure minimizes
errors due to pointing drifts. The OTF observations
were made with the 500 MHz spectrometer only.  Before each
OTF map, the pointing was updated using the CO(2--1) line emission.

\section{Results}

The individual observations of the SiO(6--5) line were examined 
for bad baselines, co-added, and multiplied by the main beam efficiency.
The resulting total SiO(6--5) line profile is shown in Fig. ~\ref{fig1} (where
it is compared with the high resolution CO(2--1) line profile, see
below).  The observed properties: the integrated 
intensity in $\rm K \times km~s^{-1}$; the peak line temperature; the
wind outflow speed $\rm V_o$; and the central velocity $\rm V_c$,
are listed in Table 1.  The central velocity and outflow velocity are in good 
agreement with the values found by S92 for the $\rm ^{13}CO$(1--0) line.
The SiO(6--5) line, like the $\rm ^{13}CO(1-0)$ line observed by S92,
is roughly parabolic in shape, although the signal-to-noise ratio is
not sufficient to definitively rule out a more complicated shape, such as
the presence of horns.  The profile does, however, show that the outflow
velocity of the dense gas in the inner envelope is about 13 $\rm 
km~s^{-1}$.

\begin{figure}
\resizebox{\hsize}{!}{\includegraphics{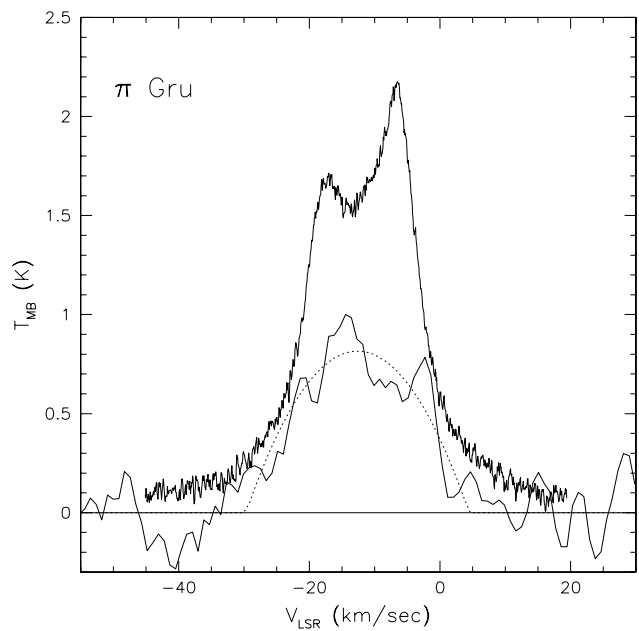}}
\caption{CO(2--1) and SiO(6--5) line profiles from the
circumstellar envelope of $\pi^1$ Gru.  Dark line: CO(2--1) line
profile observed at a resolution of $\rm 0.2 ~ km~s^{-1}$.  Light
line: SiO(6--5) line observed at a resolution of 2 $\rm km~s^{-1}$.
Dotted line: least-squares parabolic fit to the SiO(6--5) line profile.
The temperature scale of the SiO(6--5) line has been multiplied by 20.}
\label{fig1}
\end{figure}

Fig. \ref{fig1} also shows the high resolution CO(2--1) line profile.  The
shape of this line is quite different from that of the SiO(6--5) 
line profile: it has two horns and a non-parabolic Voigt profile
shape with a larger
velocity extent.
The mean velocity of the horns is -11.8 $\rm 
km~s^{-1}$, in good agreement with the central velocity of the 
SiO(6--5) and $\rm ^{13}CO$(1--0) lines.  The horns are offset in velocity 
from the line center by $\rm \pm 5.1 ~ km~s^{-1}$, about half the
outflow velocity given by the SiO(6--5) line.

\begin{figure}
\resizebox{\hsize}{!}{\includegraphics{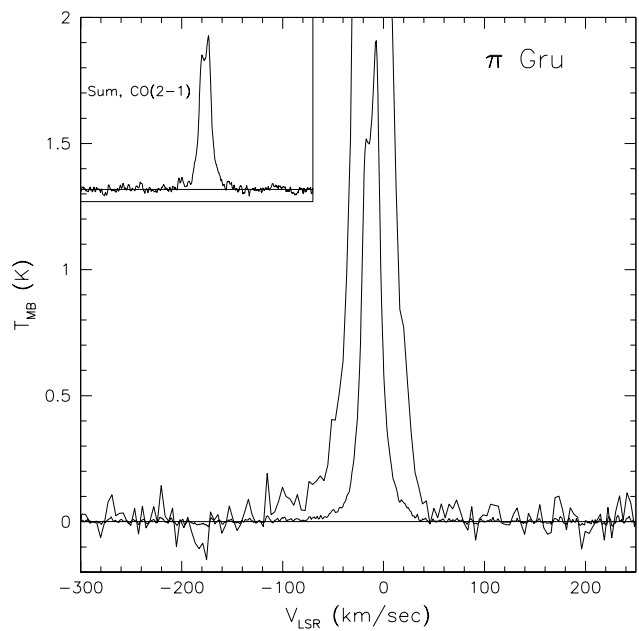}}
\caption{CO(2--1) line profile of $\pi^1$ Gru (heavy line).
Light line: scale increased by a factor of 10. The expanded profile has also
been binned to a resolution of $\rm 4 ~ km~s^{-1}$.  Insert:
the average CO(2--1) line profile from mapping observations over
$\rm \pm 30''$ about the star's position (see Fig. ~\ref{fig3}).  This
profile is plotted on a velocity range of -200 to +150 $\rm km~s^{-1}$.}
\label{fig2}
\end{figure}

Fig. ~\ref{fig2} shows the broad-band CO(2--1) line profile.  High velocity
molecular line emission is seen, extending from about -100
$\rm km~s^{-1}$ to at least +40 $\rm km~s^{-1}$, and perhaps to +80
$\rm km~s^{-1}$. This velocity range corresponds to  -88 $\rm km~s^{-1}$
to +52 $\rm km~s^{-1}$ or + 92 $\rm km~s^{-1}$
with respect to the central velocity.
The total extent of $\rm 140  - 180 ~ km~s^{-1}$
is twice that seen in the lower signal to noise ratio observations
of S92, and is almost certainly a lower limit to the true velocity extent of
the fast wind.

\begin{figure}
\resizebox{\hsize}{!}{\includegraphics{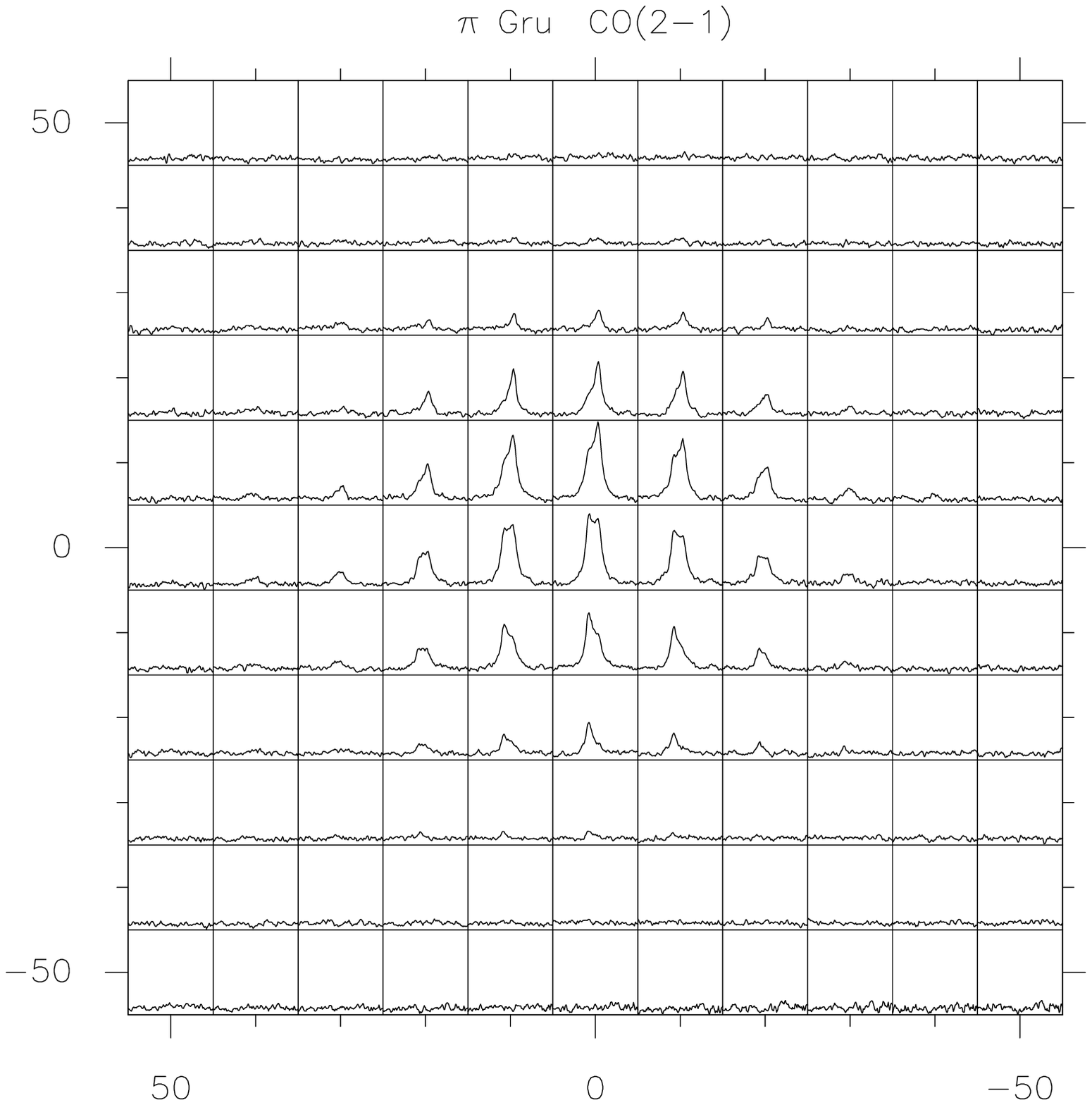}}
\caption[fig3.ps]{RA-declination
map of the CO(2--1) emission from $\rm \pi^1$ Gru.
The offsets are in arcseconds
with respect to $\alpha$(1950) = $\rm 22^h ~ 19^m ~
41.13^s$, $\delta$(1950) = $\rm -46^o ~ 12' ~ 02.4''$. The temperature scale
for each profile runs from -0.2 to 2 K and the velocity scale from -100 to
+80 $\rm km~s^{-1}$.}
\label{fig3}
\end{figure}

\begin{figure}
\resizebox{\hsize}{!}{\includegraphics{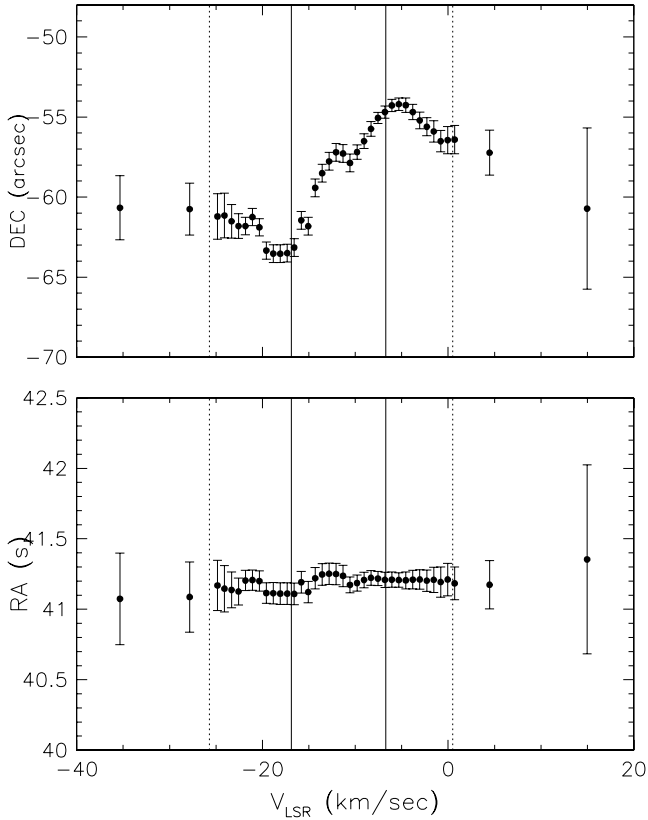}}
\caption{Dependence of the position of peak CO(2--1)
emission on velocity (a) upper panel: dependence on declination.
The ordinate is arcseconds with respect to $\rm \delta_o ~
= ~ -45^o ~ 11' ~ 00''$.  The error bars are 1$\sigma$.  The vertical solid
lines show the velocities of the horns on the CO(2--1) line observed at
the star's position (Fig. ~\ref{fig1}). The vertical dotted lines show the 
velocity extent of the SiO(6--5) emission. (b) lower panel.  As (a), for 
right ascension.  The ordinate is with respect to $\rm \alpha_o ~
= ~ 22^h 19^m ~ 00^s$.}
\label{fig4}
\end{figure}

Fig. \ref{fig3} shows the CO(2--1) line profiles in a region $\rm \pm 50''$
with respect to the star's position.  The emission is only slightly
resolved by the $\rm 30''$ beam, but kinematic structure is clearly seen.
This structure was examined by constructing a set of 11 $\times$ 11 pixel
($\rm 10''$ spacing) channel maps, i.e. maps of the emission at each
velocity. The position of the emission peak in each map was measured using the
routine IMFIT from the National Radio Astronomy Observatory's AIPS
data reduction package to fit a two-dimensional elliptical gaussian model.
The results are shown in Fig. ~\ref{fig4}, which shows the declination and right
ascension of the emission peak at each velocity.  Near the systemic velocity,
where the emission is strong, the data are plotted for every channel.
At higher relative velocities, the
data are averaged over 10 - 15 velocity channels.
Also indicated in Figure 4 are the velocities of the emission horns determined
from the central profile (Fig. \ref{fig1}) 
and the velocity range ($\rm V_c \pm V_o$)
of the SiO emission (Fig. \ref{fig1}).

Fig. ~\ref{fig4} reveals no resolvable kinematic structure in the EW
direction, but shows velocity gradients in the NS direction, in 
agreement with S92.  Fig. ~\ref{fig4} shows that
the two horns on the CO(2--1) line profile 
arise from different locations in the sky, separated by about
$\rm 10''$.  

At higher velocities with respect to the line center, $\rm \pm (6 ~ to ~
12) ~ km~s^{-1}$, the declination - 
velocity curve turns over.  
At higher velocities yet, the position 
offsets are smaller still and decrease essentially to zero.
There is a  second, lower amplitude inflection point in the
velocity - right ascension curve near the central velocity, 
with a velocity gradient of
$\rm \sim 5 ~ km~s^{-1}$ over a region a few arcseconds in diameter.

\section{A Model of the $\rm \pi^1$ Gru Envelope}

First we use a simple model for the molecular envelope to
estimate an approximate 
mass loss rate.  The SiO(6--5) line profile suggests
simple spherical outflow, and we calculated a model CO 
emission profile assuming spherical outflow at constant velocity
and mass loss rate, with a relative 
abundance f = $\rm CO/H_2 ~ = ~ 6.5 \times 10^{-4}$ (Lambert
et al. \cite{lambert}), $\rm V_o ~ = ~ 10 ~ km~s^{-1}$, and 
$\rm \mathaccent
95 M ~ = ~ 4.5 \times 10^{-7} ~ M_{\odot} ~ yr^{-1}$.  The 
model CO(2--1) line profile is parabolic and does not reproduce the
observed line shape.

We then investigated a model consisting of an expanding flattened system,
or disk, tilted to the line of sight.
The line profiles produced by 
this configuration were calculated from a model of the envelope
geometry, with the line formation calculated by a Monte Carlo radiative
transfer code based on that of Bernes (\cite{bernes}) modified for axisymmetric 
geometries.  The envelope is modeled by a set of concentric rings, 
with sufficiently fine radial spacing to approximate
a smooth distribution in optical depth (see Crosas \& Menten \cite{crosas}).
The rings have 30 radial and 30 height spacings.  Within each of the 
30 x 30 rings, the density and temperature are constant.  The 3D components
of the velocity vector are calculated at each photon step (cf. Bernes
1979).  
The level populations are solved for 30 rotational levels and for infrared
pumping to the first vibrational state (v = 0$\rightarrow$1 at 4.6 $\rm
\mu m$) using the observed 4.6 $\rm \mu m$ flux density of $\rm \pi^1$
Gru (Gezari et al. \cite{gezari}). 
The mass loss rate, radial expansion velocity, turbulent velocity, kinetic
temperature distribution and disk radius are input parameters.  The line
profile along a given line of sight is calculated by rotating the structure,
calculating the line emission across the structure and
convolving the emission
with a two-dimensional circular gaussian beam.  The fast
molecular wind was not included in this model.

\begin{figure}
\resizebox{\hsize}{!}{\includegraphics{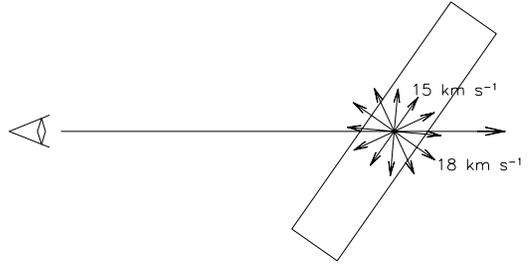}}
\caption{Proposed model of the $\rm \pi^1$ Gru circumstellar shell.
The thickness of the model disk is one-fifth of its radius.
The disk is inclined at $\rm 55^o$ to the line of sight.
North is up in this figure, east-west is perpendicular to the 
plane of the page.
This model does not include the fast wind (see text)}
\label{fig5}
\end{figure}

\begin{figure}
\resizebox{\hsize}{!}{\includegraphics{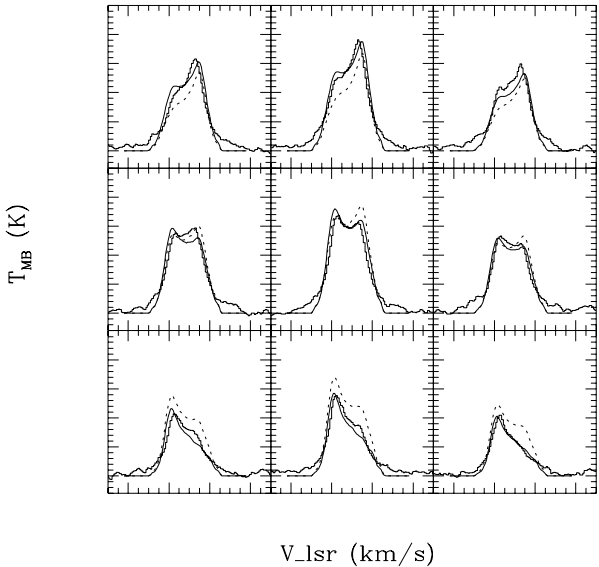}}
\caption{Comparison of observed and model CO(2--1) line profiles for the 
envelope of $\rm \pi^1$ Gru.  The geometry of the model is shown in
Fig. ~\ref{fig6} and described in the text, and the calculated line
profiles are compared with the data for the inner $\rm 20''$ of the
envelope (cf. Fig. ~\ref{fig3}).  The solid line shows the best fit
model, which includes a pointing offset of $\rm +5''$ in declination.
The dashed lines show the calculated line profile assuming that the map
is centered on the center of the envelope (see text).}
\label{fig6}
\end{figure}

A model in which the bulk of the CO(2--1) emission arises from a tilted, 
expanding disk produced by constant mass loss fits the data
well.  This model is sketched in Fig. \ref{fig5} and the model and
observed line profiles are compared in Fig. \ref{fig6}.  The northern part of 
the disk is tilted away from the observer at an angle of $\rm 55^o$ to the
line of sight.  
The temperature profile is approximated as a power law: T(r) = 
300 K $(r/10^{15} ~ cm)^{-0.7}$, as found in detailed models of AGB star
winds (Goldreich \& Scoville (\cite{goldreich}); Kwan \& Linke (\cite
{kwan}). 

The best fit model has a disk radius of $\rm 5 \times 10^{16}$ cm, a 
thickness of $\rm 10^{16}$ cm, and is produced by a constant mass loss rate 
of $\rm 1.2 \times 10^{-6} ~ M_{\odot} ~ yr^{-1}$.  The radial expansion
velocity in the plane of the disk is 15 $\rm km~s^{-1}$, increasing
slightly to 18 $\rm km~s^{-1}$ at the poles, and the turbulent velocity 
is 1 $\rm km~s^{-1}$.
It was also found that better agreement with the data is
obtained if the telescope pointing is offset from the center of the
envelope by $\rm +5''$ in declination (note that declination and 
zenith angle offsets are approximately equivalent for observations from
Hawai`i of objects which lie at very southern declinations).  The outflow
velocity in the disk agrees well with the value of $\rm 13 \pm 2 ~ 
km~s^{-1}$ observed for gas in the dense inner regions as 
observed in the SiO(6--5) line (Table 1).

\section{Discussion}

The model shown in Figures 5 and 6, of an expanding disk tilted in the
north-south direction, reproduces the observations well, and also agrees
with the higher spatial resolution observations of S92, which show that the 
emission from the envelope is elliptical with the major axis lying 
east-west.  The difference between the discussion by S92 and that here
is one of interpretation; S92 suggests that the spatially separated horn
features arise from a bipolar flow perpendicular to the disk, while our
model identifies them with the northern and southern halves of the tilted disk
(cf. Figure 5) whose projected major axis lies east-west.
While the fast wind from $\rm \pi^1$ Gru is not explicitly modeled,
the observations suggest that it is likely to be a continuation of
the velocity increase towards the poles.

The model disk which reproduces most of the emission from the inner envelope
is much smaller than the envelope extent of $\rm \sim 6 \times 10^{17}$
cm observed via 60 $\rm \mu m$ emission from dust (Young, Phillips and
Knapp 1993), and the mass loss rate required to produce the observed emission
from the disk is much higher ($\rm 1.2 \times 10^{-6} ~ 
M_{\odot}~yr^{-1}$) than the mean value ($\rm \sim 4 \times 10^{-7}
~ M_{\odot} ~ yr^{-1}$) given by the CO line intensity and 60 $\rm
\mu m$ flux density.  The envelope may thus be roughly
spherical, with a strong density increase towards the plane of the disk;
alternatively, the mass loss rate of $\rm \pi^1$ Gru may have undergone a 
large increase in the last 1000 years.

\begin{figure}
\resizebox{\hsize}{!}{\includegraphics{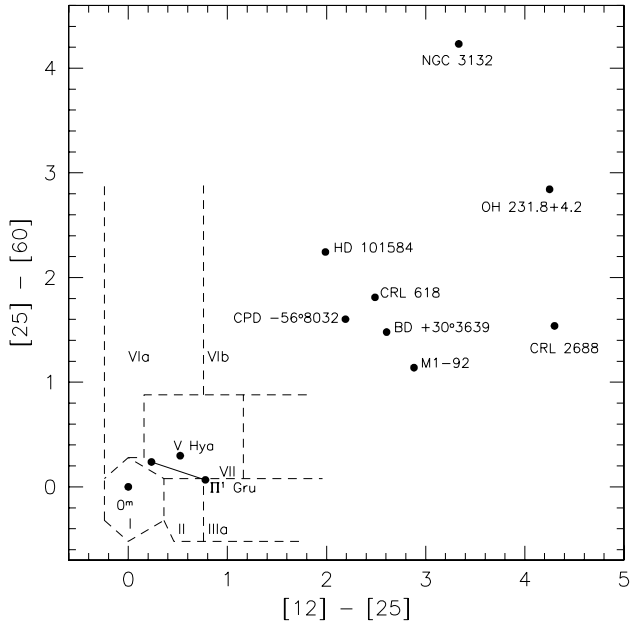}}
\caption{IRAS color-color diagram for evolved stars with fast
molecular winds (cf. Knapp et al. \cite{knapp}). 
The dashed lines outline regions 
I, II, III and VII which are
occupied by AGB stars.  The two connected points for $\rm \pi^1$ Gru correspond
to data taken at different epochs (Jorissen \& Knapp \cite{jk}).}
\label{fig7}
\end{figure}

The model proposed here for
the envelope of $\rm \pi^1$ Gru is similar to that suggested for the
envelope of the carbon star V Hya by Knapp et al. (\cite{knapp}). 
The molecular  
line emission from both envelopes is similar: the lines 
emitted from the inner envelope 
have simple parabolic shapes, the CO lines have complex, double horned 
shapes, and the velocity separation of the horns is significantly
smaller than the velocity range of the ``high density'' emission.
Also, both stars have 
fast ($\rm V_o > 50 ~ km~s^{-1}$) molecular winds.  Fig. ~\ref{fig7}
reproduces the IRAS color-color diagram for evolved stars with fast molecular
winds from Knapp et al. (1997) with the data for $\rm \pi^1$ Gru added
Jorissen \& Knapp (\cite{jk}); the
colors are plotted for the two epochs at which the star was observed.  Like
V Hya, and unlike all the other stars with fast molecular winds,
$\rm \pi^1$ Gru still has the infrared colors of an AGB star.

The structure of the $\rm \pi^1$ Gru envelope is
remiscent of structure seen in some planetary nebulae.  
For example, Manchado et al. (\cite{manchado}) 
have pointed out the existence of a class of quadrupolar
planetary nebulae, which they attribute to the presence of two bipolar outflows 
ejected in different directions.  Their proposed model for these flows (see
also Livio \& Pringle \cite{livio}; 
Guerrero and Manchado
\cite{guerrero}) is that the bipolar flow
is intermittent and that the flow axis precesses due to the precession of a 
collimating disk.  The similarity of this structure to that suggested from
these observations of the $\rm \pi^1$ Gru envelope shows that this 
complex structure is already present in the pre-existing circumstellar
envelope.  Detailed observations of the molecular emission on angular
scales of $\rm \leq 1''$ will become possible before too long; it will
be interesting to investigate the details of the complex ways in which
stars end their lives, which are only dimly discernible in the present
observations.

\section{Conclusions}

\begin{enumerate}

\item
This paper describes molecular line observations of the envelope
around the S star $\rm \pi^1$ Gru.  We find that the envelope contains
complex kinematic structure:

\begin{enumerate}

\item A fast molecular wind, with a maximum outflow
speed of at least 
70 $\rm km~s^{-1}$ and perhaps as high as 90 $\rm km~s^{-1}$.

\item A flattened, tilted disk expanding at 15 - 18 $\rm km~s^{-1}$
with a mass loss rate of $\rm 1.2 \times 10^{-6}$ $\rm M_{\odot}~yr^{-1}$
and a kinematic lifetime a factor of 10 smaller than that found for
the circumstellar dust distribution.

\end{enumerate}

\item
Our model of the molecular line emission from the $\rm \pi^1$ Gru envelope
shows that the bulk of the emission arises from a disk (and this model
also provides a good description of the envelope of the carbon star V
Hya).  It is of interest that the scale size of the velocity perturbation 
in the inner envelope (Figure 4) is
similar to the projected distance of $\rm \pi^1$ Gru's companion, since
the separation of the stars is large enough that the companion is expected
to have very little effect on a freely expanding wind (cf. the models of 
Mastrodemos \& Morris 1998; see also Jorissen 1998).   

\item
The mass loss rate required to produce the flattened disk is several times
higher than the mean mass loss rate inferred from the global CO profile
or the 60 $\rm \mu m$ emission.  The mass loss rate of $\rm \pi^1$
Gru may therefore have increased significantly in about the last 1000
years (the kinematic age of the disk).  The star may be close to the end
of its evolution on the AGB.  If so, the present observations show that 
the production of a fast molecular wind, and of complex envelope structure,
begins while the star is still on the AGB.

\end{enumerate}

\begin{acknowledgements}

We thank the CSO for providing the observing time for these observations,
the staff for their support, and Robert Lupton for help with the 
plotting program SM. We are grateful to the referee, Garrelt Mellema, for
his prompt and very useful report. This research made use of the SIMBAD
data base, operated at CDS, Strasbourg, France. Astronomical research at
the CSO is supported by the National Science Foundation via grant 
AST96-15025.  Support for this work from Princeton University
and from the N.S.F. via grant AST96-18503 is gratefully acknowledged.
\end{acknowledgements}

\end{document}